\documentclass[aps,prl,twocolumn,floatfix,superscriptaddress]{revtex4-2}
\usepackage{chemformula} 
\usepackage{siunitx}
\usepackage{mathtools}
\DeclareMathAlphabet{\altmathcal}{OMS}{cmsy}{m}{n}
\usepackage{mathptmx}
\usepackage{amsmath,amssymb}
\usepackage{color,soul}
\usepackage{ stmaryrd }
\usepackage{graphicx}


\begin{document} 

\def \ropt{$\mathrm{\rho_{opt}}$}
\def \mose2{MoSe$_2$}
\def \wse2{WSe$_2$}
\def \hBN{hBN}
\def \vbias{$V\mathrm{_{b}}$}
\def \rgv{$\mathrm{\rho_{nr}}$}
\def \rg{$\mathrm{\rho_{r}}$}
\mathchardef\mhyphen="2D

\title{Energy transfer from tunneling electrons to excitons}

\author{Sotirios Papadopoulos}
\thanks{These two authors contributed equally}
\author{Lujun Wang}
\thanks{These two authors contributed equally}
\affiliation{Photonics Laboratory, ETH Zurich, 8093 Zurich, Switzerland.}

\author{Takashi Taniguchi}
\affiliation {International Center for Materials Nanoarchitectonics, National Institute for Materials Science, 1-1 Namiki, Tsukuba 305-0044, Japan}
\author{Kenji Watanabe}
\affiliation {Research Center for Functional Materials, National Institute for Materials Science, 1-1 Namiki, Tsukuba 305-0044, Japan}

\author{Lukas Novotny}
\email{lnovotny@ethz.ch}
\affiliation{Photonics Laboratory, ETH Zurich, 8093 Zurich, Switzerland.}

\begin{abstract}
Excitons in optoelectronic devices have been generated through optical excitation, external carrier injection, or employing pre-existing charges. Here, we reveal a new way to electrically generate excitons in transition metal dichalcogenides (TMDs). The TMD is placed on top of a gold-hBN-graphene tunnel junction, outside of the tunneling pathway. This electrically driven device features a photoemission spectrum with a distinct peak at the exciton energy of the TMD. We interpret this observation as exciton generation by energy transfer from tunneling electrons, which is further supported by a theoretical model based on inelastic electron tunneling. Our findings introduce a new paradigm for exciton creation in van der Waals heterostructures and provide inspiration for a new class of optoelectronic devices in which the optically active material is separated from the electrical pathway.
\end{abstract}

\maketitle

\subsection*{Exciton generation in TMDs}

Transition metal dichalcogenides (TMDs) have attracted attention due to their interesting electronic and optical properties. In monolayer form, they exhibit a direct bandgap \cite{Mak2010}, responsible for radiative electron-hole recombination \cite{Splendiani2010}. TMDs support excitons with high binding energies due to quantum confinement \cite{Chernikov2014}, making their optical properties to be dominantly defined by their excitonic states \cite{Jung2018, Hsu2019, Ermolaev2020}. Exciton generation in TMDs was first studied through optical excitation \cite{Mak2010,Kozawa2014,Splendiani2010,Lien2019,Amani2015,Kim2021}. Later, combination with conductors (i.e. graphene) and insulators (i.e. hexagonal boron nitride) \cite{Novoselov2005,Geim2013} allowed the design and fabrication of light-emitting van der Waals (vdW) heterostructures with TMDs as the optically active material \cite{Withers2015,Paur2019,PenaRoman2020,Binder2017,Palacios-Berraquero2016,Jauregui2019,Sundaram2013,Ross2014,Uddin2022,Withers2015a,Fu2021,Binder2019,Clark2016, Lopez2022,Roman2022,Feng2022}. In these studies, excitons are electrically generated through direct charge injection \cite{Withers2015,Paur2019,PenaRoman2020,Binder2017, Palacios-Berraquero2016,Jauregui2019,Sundaram2013,Ross2014,Uddin2022,Withers2015a,Fu2021,Binder2019,Clark2016,Lopez2022,Roman2022} or through charge impact at high alternating voltages \cite{Feng2022}. Here, we introduce a different way to generate excitons in TMD based tunneling devices. We use the rich platform of 2D materials to demonstrate excitonic light emission from tunnel junctions where the TMD is located outside of the current pathway. 

\begin{figure*}
	\centering
	\includegraphics[width=0.7\textwidth]{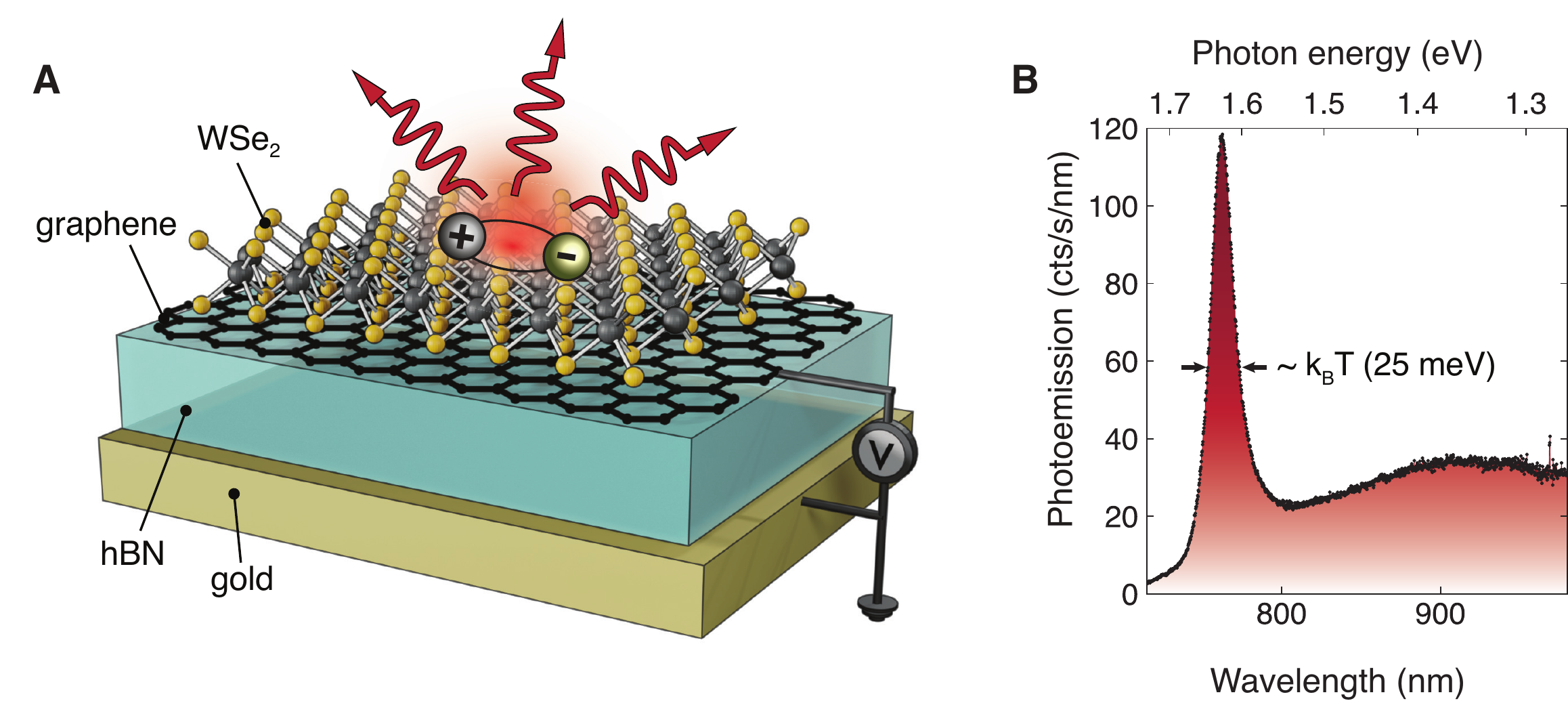}
	\caption{\textbf{Exciton generation in TMD-coupled tunnel junctions.} (a) Device composed of a \wse2-coupled graphene-hexagonal boron nitride (\hBN)-gold tunnel junction. The \wse2 layer is stacked on top of the junction. When a bias voltage \vbias~ is applied between graphene and gold, excitons are generated in \wse2. (b) Measured spectrum at room temperature of the electrically generated emission for \vbias~=~\SI{2}{V}. The peak position matches the \wse2 exciton energy and the spectral width matches the thermal energy $k\mathrm{_BT}$.} 
	\label{fig1}
\end{figure*}

The studied device structure is illustrated in Fig.~\ref{fig1}A. The TMD and the tunnel junction are electrically decoupled, preventing exciton generation through direct electron-hole injection. The emission spectrum of such a device features a distinctive peak at the exciton energy, as shown in Fig.~\ref{fig1}B. We attribute the generation of excitons to energy transfer (ET) from tunneling electrons. ET from a donor to an acceptor is mediated by dipole-dipole coupling \cite{Forster1964,Novotny2012}. It is an electromagnetic effect that is dominant at the near-field region of an emitter and has been extensively studied in biological systems \cite{Weiss1999}, molecular assemblies \cite{Dale1976}, solid-state quantum dots \cite{Kagan1996} and photosynthetic membranes \cite{van2000photosynthetic}. It has been also used for color conversion and broadband source design as an energy exchange channel between different fluorophores \cite{Achermann2006,Pimputkar2009}. Recent STM studies discuss the possibility of ET in tunneling systems \cite{Lopez2022,Roman2022} without however providing any conclusive observation. Here we make use of atomically thin 2D materials and the flat interfaces between them to preserve optical coupling in near-field distances between a TMD and a tunnel junction. Our results reveal the role of ET between electrons and excitons in tunneling-based light emitting devices.

\subsection*{TMD-coupled tunnel junctions}

Our devices are fabricated by stacking mechanically exfoliated flakes of 2D materials with a dry pickup and transfer technique (see supplementary materials for details). The resulting tunnel junctions consist of a graphene and a gold electrode separated by a hexagonal boron nitride layer (\hBN) of \SI{2.3}{nm} thickness acting as a tunneling barrier. The TMD is stacked on top of the junction  and optionally separated from the graphene electrode by a second hBN layer. 

\begin{figure*}
	\centering
	\includegraphics[width=0.7\textwidth]{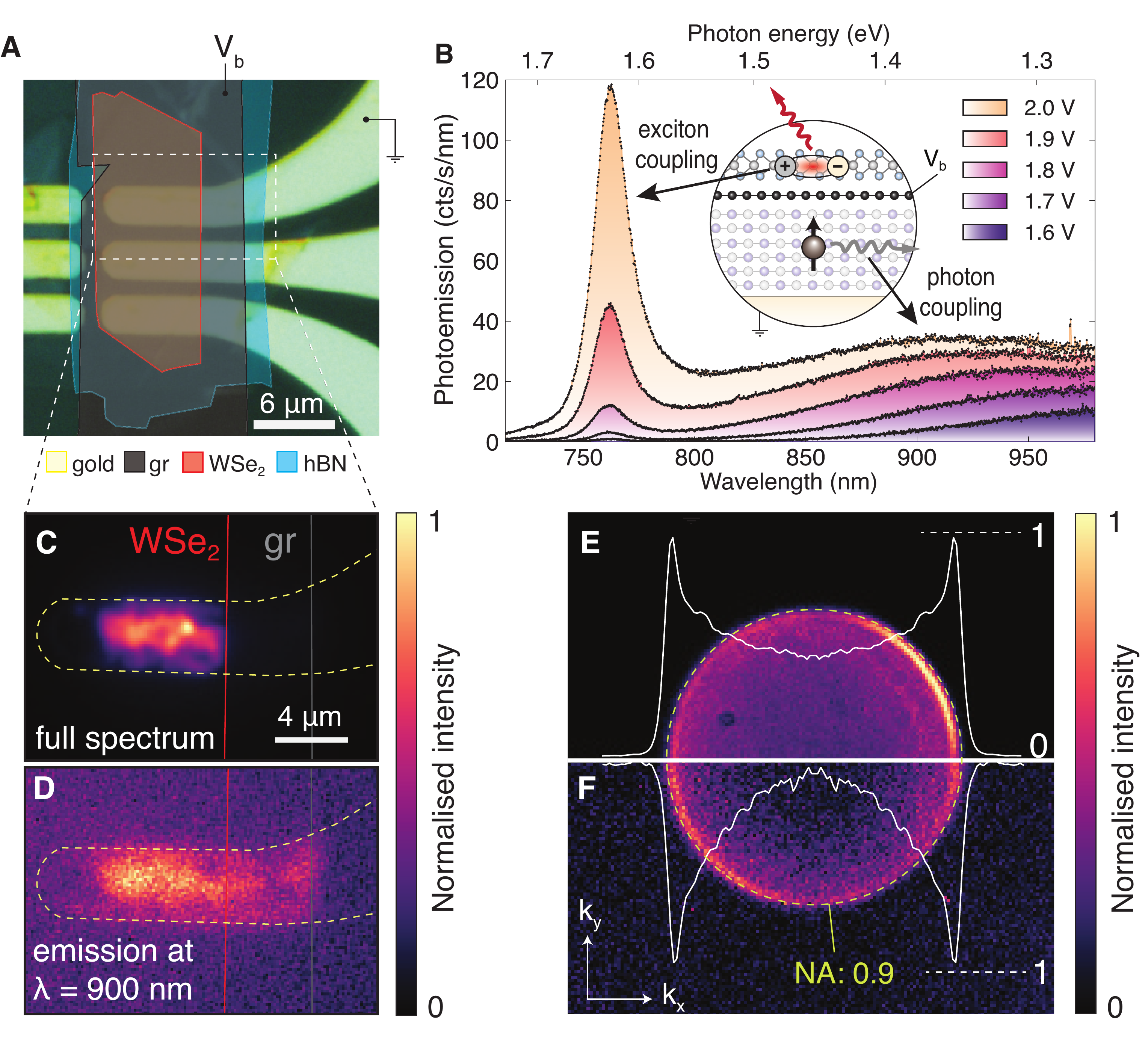}
	\caption{\textbf{ \wse2-coupled tunnel junction.} (a) Optical microscope image in false-color of a fabricated \wse2-Graphene-\hBN-gold device. The different flakes are outlined. The encapsulating~\hBN~flake is not shown for clarity. The dashed frame indicates the device area depicted in (c) and (d). (b) Recorded emission spectra from the whole device area. Contributions from radiative mode coupling (\textit{photon-coupling}) and non-radiative mode coupling (\textit{exciton-coupling}) are observed. Real space (c,d) and Fourier space (e,f) images of the emission from the dashed box in (a) for \vbias~=~\SI{2}{V}. The full emission spectrum is used in (c) and (e) whereas a bandpass filter with center wavelength 900 nm and 40 nm FWHM is used in (d) and (f). Only half of every Fourier space is imaged for ease of comparison. The white solid lines are the rotational averages of the two Fourier spaces in (e) and (f).}
	\label{fig2}
\end{figure*}

Figure~\ref{fig2}A shows a false-color microscope image of the device with an outline of the different flakes. The device area is indicated by a white dashed box. The tunnel junction area is defined by the overlap of the gold electrode and the graphene flake. The \wse2 flake is only partially covering the tunneling area. This creates two different tunneling regions, with and without \wse2 on top. We electrically and optically characterize the device by applying a bias voltage (\vbias) between the gold and graphene electrodes, as indicated in Fig.~\ref{fig2}A, and we collect the emitted light with a high-numerical aperture objective (see supplementary materials). The current and conductance measurements suggest that tunneling occurs between graphene and gold (see Fig. S1A). The spectra emitted from the whole tunneling area for various \vbias~values are presented in Fig.~\ref{fig2}B. There are two distinct contributions in the spectra. A broad emission at lower energies and a narrow emission peak at \SI{1.63}{eV} (\SI{762}{nm}). The broad emission shifts to higher photon energies as \vbias~is increased. This behavior suggests that the emission originates from inelastic electron tunneling that couples to radiative modes of the free-space continuum (\textit{photon-coupling}) with e\vbias~being the cut-off energy of emission \cite{Parzefall2015,Parzefall2019}. The narrow peak at \SI{1.63}{eV} corresponds to the neutral 1s exciton of \wse2 and its narrow width of around \SI{25}{meV} can be attributed to graphene filtering \cite{Lorchat2020}. It has been shown that charge transfer between graphene and TMDs, filters the TMD photoluminescence spectra from any charged exciton contributions \cite{Lorchat2020}. To further our understanding, we study the emission distribution for different spectral regions. Real space images of the photoemission are presented in Fig.~\ref{fig2}C, where we observe that the area covered by \wse2 dominates the emission and in Fig.~\ref{fig2}D, where the emission is filtered by a band-pass filter at \SI{900}{nm}. This is a spectral region where the excitonic contribution is much reduced and the emission extends over both device areas. This indicates that the broad low energy contribution indeed originates from \textit{photon-coupling} and is not associated with the presence of \wse2. To understand the angular emission of the device  we record Fourier-space images. Fig.~\ref{fig2}E shows the Fourier space image of the full emission spectrum whereas Fig.~\ref{fig2}F depicts the Fourier space image of the bandpass filtered light around \SI{900}{nm}. For the spectrally filtered measurement at 900 nm, the emission vanishes at the center of the Fourier space ($k_x=k_y=0$) (Fig.~\ref{fig2}F) indicative for an out-of-plane dipole orientation. This is characteristic of inelastic electron tunneling where the transition dipole is oriented along the electron path. On the other hand, when measuring the full spectrum, the Fourier space image (Fig.~\ref{fig2}E) is not zero at $k_x=k_y=0$, which indicates an in-plane contribution. This extra contribution is associated with the excitonic emission of \wse2, which is known to originate from in-plane transitions \cite{Schuller2013}. We emphasize that in our devices any exciton generation due to electron-hole injection is avoided due to the \wse2 being outside of the tunneling pathway. This is further supported by the fact that very similar emission spectra are observed for negative \vbias~(see Fig. S2). For these reasons we conclude that excitons are generated in the TMD by ET from tunneling electrons. \\
\indent 
We continue to investigate this exciton generation process by considering the band-structure diagram shown in  Fig.~\ref{fig3}A. Upon application of \vbias, electrons tunnel from one electrode to the other, elastically or inelastically. In the latter case, the transition dipole associated with the energy loss $\Delta E$ couples to available optical modes of the environment. Some of these modes are radiative (photons) as in the case of the broad emission observed in Fig.~\ref{fig2}A, or non-radiative, as in the case of ET. This ET mechanism generates excitons in \wse2  and their spontaneous decay contributes to the narrow emission peak in the spectra of Fig. \ref{fig2}B. Thus, the excitonic emission observed in our device is a two step process, in which the transition energy is first transferred to the \wse2 exciton and then spontaneously emitted (\textit{exciton-coupling}). \\
\indent 
Next, we compare the ET efficiency to the \textit{photon-coupling} efficiency. To do that we first break down the processes to the involved interactions (see pictorial representation in Fig.~\ref{fig3}B) and we assign conversion efficiencies. Both \textit{photon-} and \textit{ exciton-coupling} processes describe an electron-to-photon ($e\mhyphen\gamma$) conversion. \textit{Photon-coupling} is a first order process for which we define an $e\mhyphen\gamma$ conversion efficiency as $\eta\mathrm{_{e\mhyphen\gamma}}$. \textit{Exciton-coupling} is a second order process in which we can assign a combined efficiency $\eta^{\prime}\mathrm{_{e\mhyphen\gamma}} = \eta\mathrm{_{e\mhyphen x}}\cdot\eta\mathrm{_{x\mhyphen\gamma}}$, where $\eta\mathrm{_{e\mhyphen x}}$ is the ET efficiency and $\eta\mathrm{_{x\mhyphen\gamma}}$ is the exciton to photon conversion efficiency. From the spectra shown in Fig.~\ref{fig2}B, we can infer a value of 4.3 for the ratio between the photon-coupled emission and the exciton-coupled emission. By taking into account the values for \wse2 PL efficiency and the graphene quenching (see supplementary materials) we arrive at an estimation of the ratio $\frac{\eta_{e\mhyphen x}}{\eta_{e\mhyphen\gamma}} \cong 10^4$. This result indicates that ET is orders of magnitude more efficient than direct photon emission. In order to shed light on this surprising finding we analyze in the next section the density of optical states near a monolayer TMD.

\subsection*{Optical density of states near monolayer TMDs}

\begin{figure*}
	\centering
	\includegraphics[width=1\textwidth]{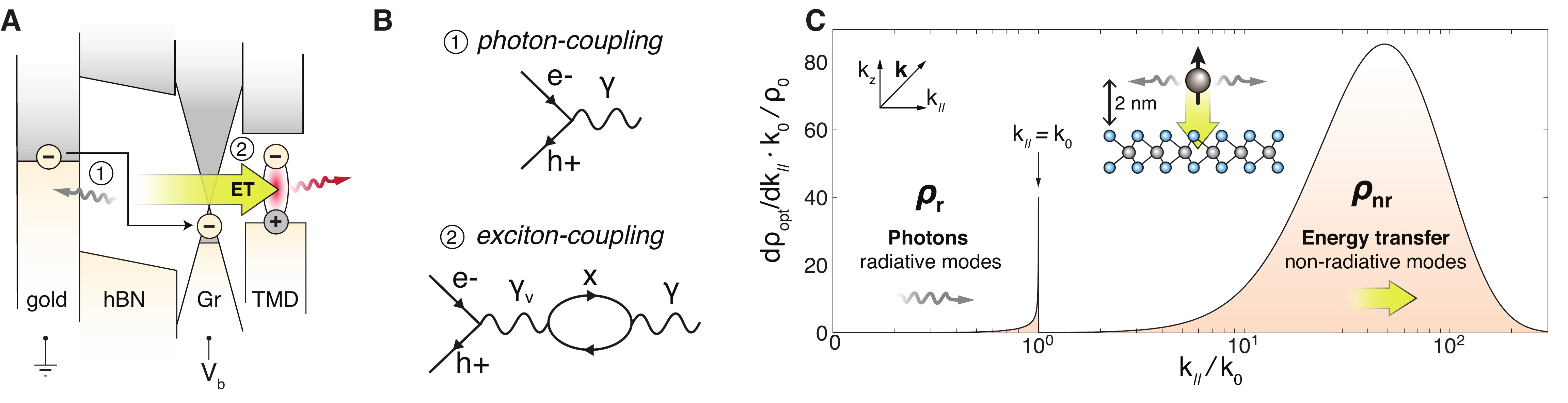}
	\caption{ \textbf{Photon and exciton coupling of inelastic tunneling electrons.} (a) Band diagram of the TMD-coupled tunnel junction. A voltage \vbias~ is applied across the \hBN~barrier between graphene and gold electrodes. Part of the electrons tunnel inelastically. They either couple to radiative modes (1) or couple non-radiatively to the TMD via ET (2). The generated excitons decay radiatively making the ET process detectable in the farfield. (b) Pictorial representation of the physical processes involved in the light emission. In (1) electron-hole recombination is followed by photon emission. In (2) ET is depicted as a virtual photon $\gamma_v$ that couples to the TMD and creates an electron-hole pair (x). (c) Angular spectral density of~\ropt~for the case of an out-of-plane dipole placed \SI{2}{\nano \meter} away from a monolayer \wse2. Optical properties are taken from \cite{Jung2018}. The contribution of non-radiative modes (\rgv) is substantially higher than the contribution of radiative modes (\rg). The calculation assumes a photon energy of \SI{1.68}{eV}, matching the exciton energy of  \cite{Jung2018}. The x-axis is normalized by the vacuum wavenumber $k_0$ and the y-axis is normalized by $\rho_0/k_0$, where $\rho_0$ is the vacuum LDOS. Inset in (c) illustrates the calculated system.} 
	\label{fig3}
\end{figure*}

When a dipole interacts with an absorbing material at a distance much smaller than the wavelength, the largest fraction of the dissipated power is associated with non-radiative energy transfer (ET) from the dipole to the material~\cite{Ford1984,Andrews2004}. The rate of this process is related to the local density of optical states (LDOS)~\ropt. In Fig. \ref{fig3}C we show the calculated angular spectral density of the LDOS, $\frac{\mathrm{d \rho\mathrm{_{opt}}} }{\mathrm{d} k\mathrm{_\parallel}}$, as a function of $k_\parallel$ for the simple case of a dipole at \SI{2}{nm} distance from a TMD monolayer. Here, $k_\parallel$ is the wavevector component along the plane of the TMD monolayer  (see inset in Fig. \ref{fig3}C) and $k\mathrm{_0}$ is the wavenumber for a photon energy equal to the exciton energy of \wse2. The calculated~\ropt~for $k_\parallel < k_0$ includes radiative modes associated with photon emission (\rg) whereas the region $k\mathrm{_\parallel} > k\mathrm{_0}$ accounts for non-radiative modes (\rgv) associated with ET. \rgv~is orders of magnitude higher than \rg. Moreover, the influence of the TMD on \rg~ enhancement is negligible. Hence, the TMD introduces mainly additional non-radiative decay channels that are associated with near-field ET ($\rho\mathrm{_{opt}} \cong \rho\mathrm{_{nr}}$), and is the reason for the high $\frac{\eta\mathrm{_{e\mhyphen x}}}{\eta\mathrm{_{e\mhyphen\gamma}}}$ ratio. The ET process in our measurements is highly efficient and shows that near-field interactions at nanoscale distances are dominating the LDOS. Here, we probe this non-radiative interaction by using a direct-gap semiconductor whose luminescence is a direct measure for the ET efficiency. In essence, the TMD acts as a receiving optical antenna that enhances the LDOS and converts non-radiative modes to a measurable signal. We refer to this process as electron-photo-luminescence (ePL) since it is triggered by an electron that transfers energy through optical modes to \wse2.

\subsection*{Dependence on TMD-Graphene separation}

\begin{figure*}
	\centering
	\includegraphics[width=0.7\textwidth]{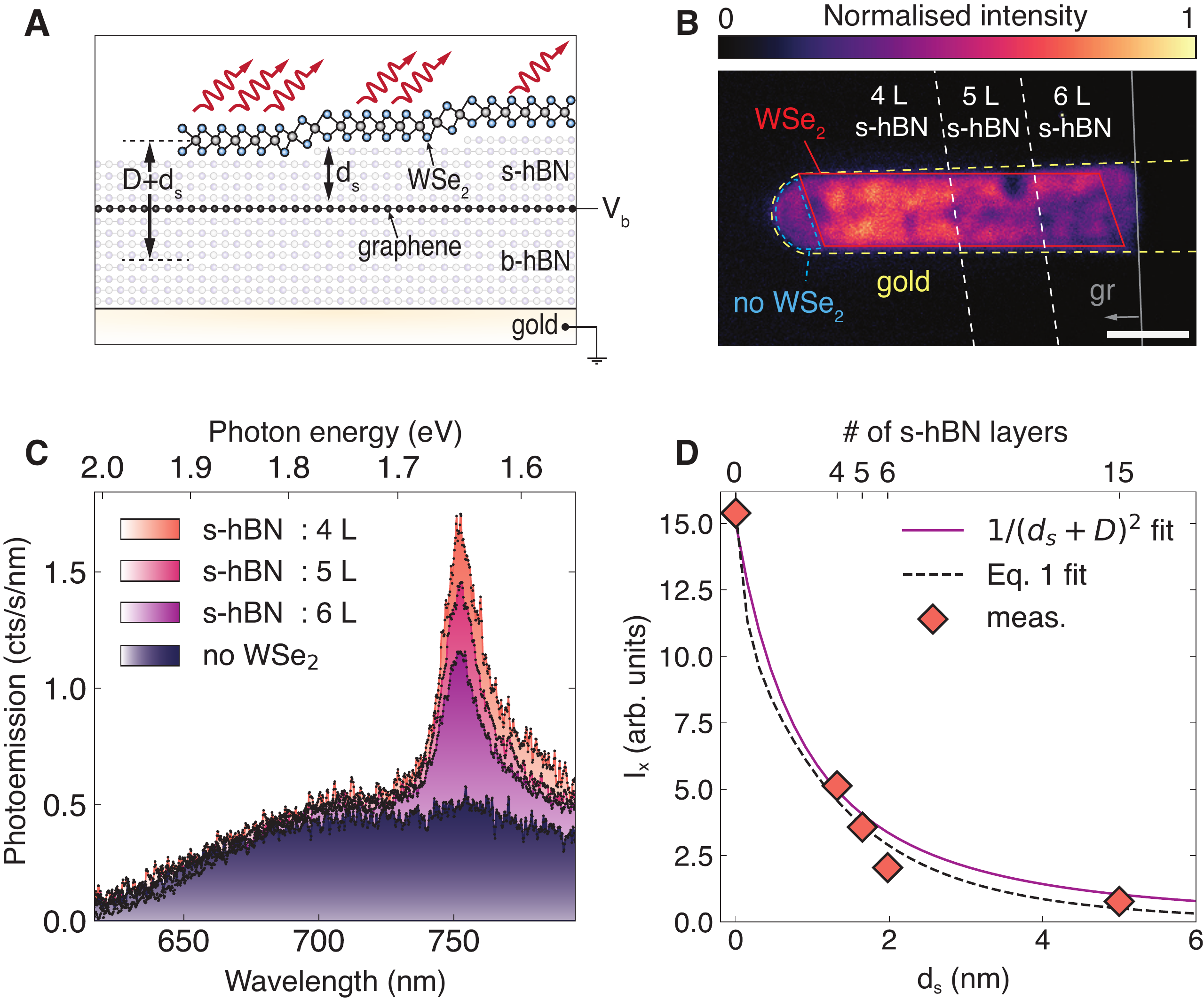}
	\caption{\textbf{Distance dependent energy transfer. }(a) Schematic of a TMD separated from the tunnel junction by a spacer-\hBN~(s-\hBN). The coupling strength between TMD and the tunnel junction is varied using different s-\hBN~thicknesses $d_\mathrm{s}$. (b) Real-space image of emission from the \wse2/s-\hBN/Gr/b-\hBN/gold device. Strongest emission is observed from the thinnest s-\hBN~region. The emission is  filtered with a bandpass filter ($\lambda_{center}$ = \SI{750}{\nano\meter} and fwhm = \SI{40}{\nano\meter}). The scale bar is \SI{4}{\micro\meter}. (c) Spectra measured on areas with different s-\hBN~thickness and for \vbias~=~\SI{2}{V}. The broad spectral background corresponds to direct photon coupling and is also observed in areas with no \wse2. (d) Excitonic emission I$_x$ as a function of $d\mathrm{_s}$. The values are normalized with the PL efficiency, the emission area and the intensity of background emission. The measurement point at $d\mathrm{_s}$ = \SI{0}{nm} refers to the device presented in Fig.~\ref{fig2}. The three points  between  $d\mathrm{_s}$ = \SI{1}{nm} and \SI{2.5}{nm} refer to the device presented in Fig.\ref{fig4}B and the point at $d\mathrm{_s}$ = \SI{5}{nm} refers to a third device, for which spectra are presented in Fig. S2. The dashed curve is the result of Eq.~\ref{eq1} and the solid curve is a fit with an inverse square distance function. $\mathrm{D+d_s}$ is the distance between the TMD and the center of the b-hBN layer as illustrated in (a).} 
	\label{fig4}
\end{figure*}

To further test our interpretation, we study the dependence of the excitonic emission on the distance between the TMD and the tunnel junction. A device schematic is shown in Fig.~\ref{fig4}A where we vary the coupling to the TMD by changing the thickness $d_\mathrm{s}$ of a \hBN~spacer layer (s-\hBN) in different device regions. Figure~\ref{fig4}B shows the emitted light from a device where the s-\hBN~ presents steps in thickness in the tunnel junction area. Four regions of emission are created. Three with different s-\hBN~ thicknesses (6, 5 and 4 layers) and one without \wse2 on top. We observe that the emission gets stronger the thinner the s-hBN is. As evidenced by the spectra shown in Fig.~\ref{fig4}C the enhanced emission can be entirely attributed to the strength of the excitonic peak. We thus find that the excitonic contribution increases for thinner s-\hBN~ in contrast to the broadband background that doesn't show any major change. A comparison of the excitonic emission intensity $I\mathrm{_x}$ for different $d\mathrm{_s}$ at \vbias~=~\SI{2}{V} is given in Fig.~\ref{fig4}D. To compare different devices we normalize $I\mathrm{_x}$ by the area of emission, the PL efficiency and the background level (see supplementary materials). The data points in Fig.~\ref{fig4}D clearly show that the rate of exciton generation, and hence the ET rate, decay superlinearly with the separation $d_\mathrm{s}$. 

$I\mathrm{_x}$ can be modeled by calculating the exciton generation rate $\Gamma\mathrm{_x}$. By using inelastic tunneling theory \cite{Parzefall2017} $\Gamma\mathrm{_x}$ assumes the following expression:

\begin{equation}
	\begin{aligned}
		\Gamma\mathrm{_x} (V\mathrm{_b},d\mathrm{_s}) \propto & \\ \int_{0}^{\infty}\eta\mathrm{_{abs}}(\omega)&\frac{\gamma\mathrm{^{0}_{inel}}(\omega,V\mathrm{_b})}{\rho\mathrm{_{0}}(\omega)} [\rho\mathrm{_{opt}}(\omega,d\mathrm{_s}) -\rho\mathrm{_{opt}}(\omega,\infty)]d\omega 
	\end{aligned}
	\label{eq1}
\end{equation}

\noindent
where $\omega$ is the angular frequency, $\eta\mathrm{_{abs}}$ is the absorption spectrum of monolayer \wse2 \cite{Kozawa2014},  $\gamma\mathrm{^{0}_{inel}}$ is the inelastic tunneling spectral rate in vacuum and $\rho\mathrm{_{0}}$ is the vacuum LDOS. We approximate the LDOS responsible for exciton generation by $\rho\mathrm{_{opt}}(\omega,d\mathrm{_s})-\rho\mathrm{_{opt}}(\omega,\infty)$ which assumes that the states provided by the TMD, leading to exciton generation, vanish for $d\mathrm{_s} \shortrightarrow \infty$. See supplementary materials for more information on the calculation. The calculated $\Gamma\mathrm{_x}$ as a function of distance $d\mathrm{_s}$ and for \vbias~=~\SI{2}{V} is plotted in Fig.~\ref{fig4}D. It agrees well with the distance dependent measurement. Interestingly, both $\Gamma\mathrm{_x}$ and the measured data can be described by a simple inverse square law, as shown in Fig.~\ref{fig4}D. Similar observations were reported for ET between optically excited TMDs \cite{Karmakar2022}. The  good agreement between our theoretical model and our experimental measurements supports our interpretation.

\subsection*{Monolayer vs bilayer TMD}

\begin{figure*}[t!]
	\centering
	\includegraphics[width=1\textwidth]{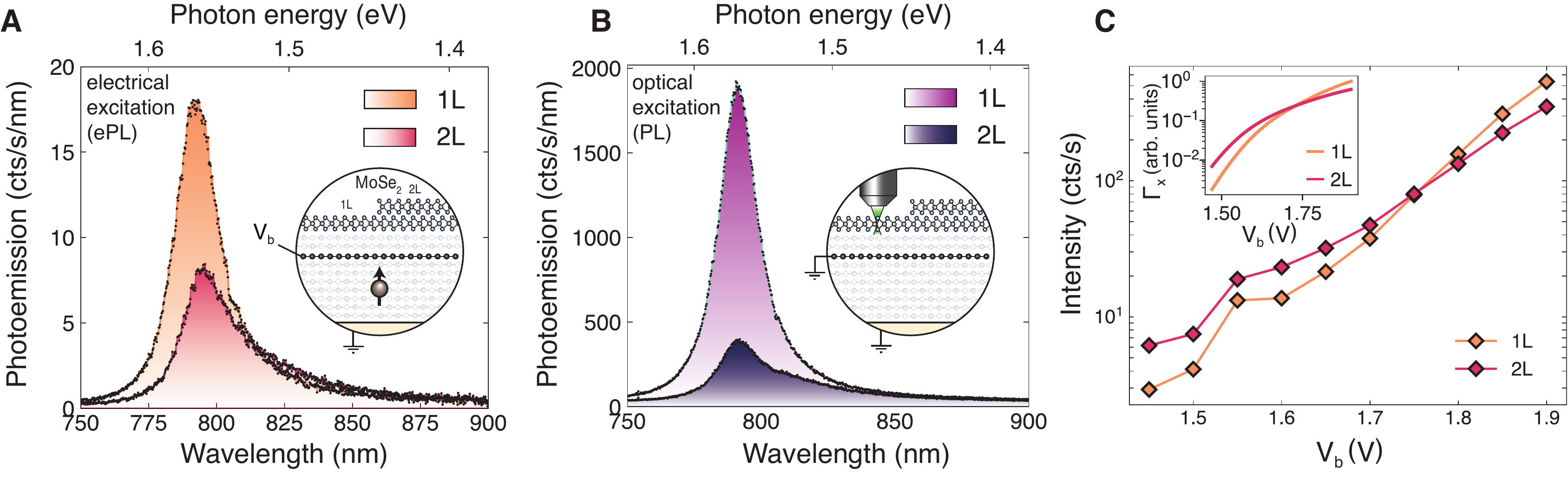}
	\caption{\textbf{Tunnel junctions coupled to monolayer and bilayer \mose2.} (a) ePL emission spectra from a device with monolayer (1L) and bilayer (2L) \mose2. Spectra taken at \vbias~=~\SI{1.9}{V}. The inset shows the device schematic. (b) PL emission from 1L and 2L regions of the same device. (c) Integrated intensities of the ePL spectra in (a) as a function of \vbias. The inset shows corresponding theoretical curves based on Eq.\ref{eq1}.} 
	\label{fig5}
\end{figure*}

We continue our study by using a different TMD (\mose2) and by comparing ET for monolayers (1L) and bilayers (2L). Figure~\ref{fig5}A shows a schematic of the device and the measured ePL spectra. The emission for 2L \mose2 ($I\mathrm{_{2L}^{ePL}}$) is less intense than for 1L \mose2 ($I\mathrm{_{1L}^{ePL}}$) owing to the indirect bandgap of 2L \mose2. To compare this ET-mediated photoluminescence (ePL) with optically excited photoluminescence (PL) we optically excite the two regions with a laser and measure the corresponding emission spectra $I\mathrm{_{1L}^{PL}}$ and $I\mathrm{_{2L}^{PL}}$, respectively. The measurements are presented in Fig.~\ref{fig5}B. The center wavelengths and the shapes of PL and ePL spectra match very well, which supports the interpretation that ePL corresponds to spontaneous exciton emission. We note however, that the ratios $I\mathrm{_{1L}^{PL}}$/$I\mathrm{_{2L}^{PL}}$ and $I\mathrm{_{1L}^{ePL}}$/$I\mathrm{_{2L}^{ePL}}$ are different. Interestingly, the $I\mathrm{_{1L}^{ePL}}$/$\mathrm{I_{2L}^{ePL}}$ ratio depends on \vbias. In fact, $I\mathrm{_{2L}^{ePL}}$ is stronger than $I\mathrm{_{1L}^{ePL}}$ for low voltages and a crossing occurs near \vbias~=~\SI{1.75}{V}, as shown in Fig.~\ref{fig5}C, which depicts the integrated spectra for different voltages. By employing the model in Eq.~\ref{eq1} we are able to reproduce this effect. The result of this calculation is given in the inset of Fig.~\ref{fig5}C. The reason for this voltage dependence are the different optical properties of 1L and 2L \mose2, i.e. the absorption spectrum $\eta\mathrm{_{abs}}$ of 2L MoSe2, which features a cut-off energy at lower energies and stronger absorption than the 1L \mose2. This explains the earlier onset-voltage for $I\mathrm{_{2L}^{ePL}}$ in Fig.~\ref{fig5}C. Moreover, the two curves are also affected by the dependence of the LDOS on flake thickness. In fact, our model predicts that the rate of ET to a 2L TMD is lower compared to a 1L despite the fact that the 2L TMD is thicker. This surprising behavior has been studied in the past in optically excited systems \cite{Prins2014}.

\subsection*{Discussion}

The exciton generation mechanism, discussed in this work, is an electromagnetic phenomenon that is usually neglected in the analysis of tunneling driven systems. The tunneling probability of an electron relies on the available electronic states but also optical states. In semiconducting materials, exciton resonances provide non-radiative optical states that couple strongly with inelastic electrons increasing the tunneling probability. This leads to near-field generation of excitons even for applied voltages below the electronic bandgap of the material, thus contributing to sub-bandgap emission. Such low voltage emission has been previously studied with interpretations varying from direct exciton formation \cite{Binder2017} to Auger scattering \cite{Binder2019}. Here we show that ET from tunneling electrons can contribute to this sub-bandgap emission.  

\subsection*{Conclusions}

We investigated ET from a tunnel junction to TMD excitons. Our studies are based on vdW tunnel devices in which the TMD is placed outside the electronic pathway, ensuring that there is no direct charge injection into the TMD. We studied photoemission spectra as well as real space and Fourier space emission patterns and we concluded that excitons are generated when the tunnel junction is under bias. This surprising observation is understood by ET from tunneling electrons to excitons. Emission from excitons is observed even when the TMD is separated from the junction by a \hBN~spacer layer. Our calculations reveal that non-radiative modes of the LDOS are strongly increased at the exciton energy. These non-radiative modes are responsible for the efficient energy transfer from the tunnel junction to the TMD. Our theoretical model based on inelastic tunneling theory and near-field coupling is in agreement with our experimental measurements. \\ \indent The electrical generation of excitons via ET provides new perspectives for the development of optical sources, detectors and sensors. For example, it can be employed for truly chip-scale light sources with applications in electrical pumping scheme for vdW heterostructure lasers \cite{Ye2015,Paik2019} or optical sensing schemes \cite{Oh2021} that do not rely on external laser sources. 

\section*{Acknowledgments}

The authors would like to thank Mathieu Luisier, Achint Jain, Martin Frimmer, Ronja Khelifa, Anna Kuzmina, Shengyu Shan and Massimiliano Rossi for fruitful discussions. This study was supported by funding from ETH Zurich under ETH Grant No. ETH-15 19-1 SYNEMA, the ETH Zurich Foundation project number 2013-08 (11) with a donation from the Stavros Niarchos Foundation, and the Swiss National Science Fund under grant number $200020\_192362$. K.W. and T.T. acknowledge support from the JSPS KAKENHI (Grant Numbers 19H05790, 20H00354 and 21H05233).

\section*{Author contributions}

S.P., L.W. and L.N. conceived the experiment. S.P. fabricated the devices, performed the measurements, analysed the data, developed the theoretical model and performed the calculations. L.W. supported in the device fabrication. L.W. and L.N. helped with the data interpretation. K.W. and T.T. provided the high-quality \hBN~crystals, L.N. initiated and supervised the project. S.P. wrote the paper and all authors discussed the results and worked on the manuscript. 

\end{document}

\renewcommand{\thesection}{S\arabic{section}}
	\renewcommand{\thefigure}{\textbf{S\arabic{figure}}}
	\def \ropt{$\mathrm{\rho_{opt}}$}
	\def \mose2{MoSe$_2$}
	\def \wse2{WSe$_2$}
	\def \hBN{hBN}
	\def \vbias{$V\mathrm{_{b}}$}
	\def \rgv{$\mathrm{\rho_{nr}}$}
	\def \rg{$\mathrm{\rho_{r}}$}
	\mathchardef\mhyphen="2D

	\title{Supplementary materials for \\ \textbf{Energy transfer from tunneling electrons to excitons}} 
	
	%\title{Near-field coupling of van der Waals tunnel junctions to excitons in transition metal dichalcogenides}
	%\title{Electrically controlled van der Waals tunnel junctions coupled to transition metal dichalcogenides}
	
	
	%\thanks{}
	
	\author{Sotirios Papadopoulos}
	\thanks{These two authors contributed equally}
	\author{Lujun Wang}
	\thanks{These two authors contributed equally}
	\affiliation{Photonics Laboratory, ETH Zurich, 8093 Zurich, Switzerland.}
	
	\author{Takashi Taniguchi}
	\affiliation {International Center for Materials Nanoarchitectonics, National Institute for Materials Science, 1-1 Namiki, Tsukuba 305-0044, Japan}
	\author{Kenji Watanabe}
	\affiliation {Research Center for Functional Materials, National Institute for Materials Science, 1-1 Namiki, Tsukuba 305-0044, Japan}
	
	\author{Lukas Novotny}
	\email{lnovotny@ethz.ch}
	\affiliation{Photonics Laboratory, ETH Zurich, 8093 Zurich, Switzerland.}

%%%%%%%%%%%%%%%%% END OF PREAMBLE %%%%%%%%%%%%%%%%

\maketitle

\subsection*{Materials and methods}
\paragraph*{\textbf{Sample fabrication}}
\mose2,  \wse2, graphene and \hBN~flakes are mechanically exfoliated from bulk crystals on Si/SiO$_2$ substrates with the scotch tape method. The \hBN~flakes are exfoliated in air the other flakes in argon (Ar) atmosphere. The Si/SiO$_2$ substrates are first cleaned in an oxygen plasma asher for 5 minutes. The heterostructures are fabricated by a dry pick-up and transfer technique \cite{Zomer2014} as follows: A poly-dimethylsiloxane (PDMS) stamp, covered by a thin polycarbonate (PC) film, is used to sequentially pick-up all flakes from their substrate at 80 °C in Ar atmosphere in a glovebox. The resulting stack is then transferred on top of pre-patterned gold electrodes on a glass substrate by heating up the substrate to 170 °C in order to melt the PC film and allow it to detach from the PDMS stamp. In that way, the stack is transferred on top of the gold electrodes. Later, the PC film is dissolved in a chloroform bath for 30 minutes. The gold electrodes are pre-patterned by photolithography followed by e-beam evaporation of 5 nm Ti (adhesion layer) and 50 nm gold. Metal lift-off is done in an acetone bath at 60 °C. \\ \\

\paragraph*{\textbf{Electrical and optical characterization}}

The gold electrodes on the sample's substrate are wire-bonded to adhesive copper pads that allow conductive contact to co-axial connections to a Keithley Source Meter 2602B, that is used as a voltage source and current meter. The sample is then put, facing down, on a Nikon TE300 inverted microscope in ambient conditions. The light emitted by the device under bias voltage is collected by a 100x objective with 0.9 NA. All spectra are measured by imaging the real space of emission at the entrance of a Princeton Instruments Acton SpectraPro 300i spectrometer. The real and Fourier space (back-focal plane) of emission is imaged at the entrance of an Andor iXon Ultra camera. Finally, all PL measurements were collected by optically exciting the sample with a 532 nm solid state laser. \\ \\

\paragraph*{\textbf{Density of optical states}}

The dissipated power $P$ of a point electric dipole is calculated by \cite{Novotny2012}

\begin{equation}
	P = \frac{1}{2}\omega~ \mathbf{p}~ Im \{\mathbf{E(r_0)}\}
\end{equation}
\noindent
where $\omega$ is the angular frequency of the dipole's emission, $\textbf{p}$ is the dipole moment and $\textbf{E}$ is the electric field at the dipoles origin \textbf{$r\mathrm{_0}$}. The dipole is assumed to be situated at the center of the tunneling \hBN~layer. The electric field  $\textbf{E(r$_0$)}$ is calculated by solving the electromagnetic wave equation for a multi-layer structure. The optical density of states \ropt~is then calculated through its relation to P \cite{Novotny2012}

\begin{equation}
	\frac{\rho\mathrm{_{opt}}}{\rho\mathrm{_{0}}} = \frac{P}{P\mathrm{_{0}}}
\end{equation}
\noindent
with $P\mathrm{_{0}}$ being the dissipated power for a point dipole in vacuum. \\ \\

\paragraph*{\textbf{Exciton to photon coupling ratio}}

We calculate the ratio $\eta\mathrm{_{e\mhyphen x}}/\eta\mathrm{_{e\mhyphen\gamma}}$ using the following relationship,

\begin{equation}
	\frac{I\mathrm{_x}}{I\mathrm{_\gamma}} =  \frac {\eta\mathrm{_{e\mhyphen x}}\cdot\eta\mathrm{_{x\mhyphen\gamma}}\cdot\eta\mathrm{_{col}^{\shortrightarrow}}}{ \eta\mathrm{_{e\mhyphen\gamma}}\cdot\eta\mathrm{_{col}^{\shortuparrow}}} 
\end{equation}
\noindent

where $I\mathrm{_x}$ is the exciton coupled emission, $I\mathrm{_\gamma}$ is the photon coupled emission, $\eta\mathrm{_{col}^{\shortrightarrow}}$ and $\eta\mathrm{_{col}^{\shortuparrow}}$ is the collection efficiency for an in-plane and an out-of-plane dipole, respectively. The $\eta\mathrm{_{x\mhyphen\gamma}}$ expresses the PL efficiency of the \wse2 flake which is measured to be around $1.5\cdot 10^{-5}$. The low value results from graphene quenching \cite{Lorchat2020}. \\ \\

\paragraph*{\textbf{Exciton generation rate}}

The exciton generation rate calculated in Eq. 1 in the main text involves the inelastic tunneling rate spectrum $\gamma\mathrm{_{inel}}$ which is a function of the electronic density of states of the right and left electrodes ($\rho\mathrm{_R}$ and $\rho\mathrm{_L}$, respectively),the optical density of states $\rho\mathrm{_{opt}}$ and the applied \vbias. It is calculated by \cite{Parzefall2019a,Parzefall2017},

\begin{equation}
	\gamma\mathrm{_{inel}}(\hslash\omega) = \frac{\rho\mathrm{_{opt}}(\hslash\omega)\pi e^2}{3\hslash\omega m^2\epsilon\mathrm{_0}}\int_{\hslash\omega}^{eV\mathrm{_b}}|\altmathcal{P}(E,\hslash\omega)|^2\rho\mathrm{_R}(E-\hslash\omega)\rho\mathrm{_L(E)}dE
	\label{eq2}
\end{equation}
\noindent
where $\altmathcal{P}$ is the momentum matrix element, \textit{m} is the electrons mass, $\epsilon_0$ is the vacuum permittivity and \textit{e} is the elementary charge. \\ \\

\paragraph*{\textbf{Emission comparison between different devices}}
To be able to compare the rate of excitonic emission of different devices we normalize the measured emission intensity $I\mathrm{_x}$ by the area of emission and by the relative PL efficiency of the flake. Finally we normalize by the photon-coupled background emission $I\mathrm{_\gamma}$ in every device. This normalization is necessary because in every device the tunneling barrier thickness may vary by $\pm \SI{0.33}{nm}$. This has a minor effect on the optical coupling but it can drastically affect  $\gamma\mathrm{_{inel}}$ through the momentum matrix element $\altmathcal{P}$ (Eq.~\ref{eq2}). This impacts the exciton generation rate $\Gamma \mathrm{_x}$ making the direct comparison between different devices impossible. Since $I\mathrm{_\gamma}$ is directly proportional to $\altmathcal{P}$ we use it to normalize the $I\mathrm{_x}$ emission and be able to compare devices with different TMD to tunnel junction distance (Fig.~4 in the main text). \\ \\

\subsection*{Conductance measurements}
\begin{center}
	\begin{figure}[h!]\center
		\includegraphics[width=0.7\textwidth]{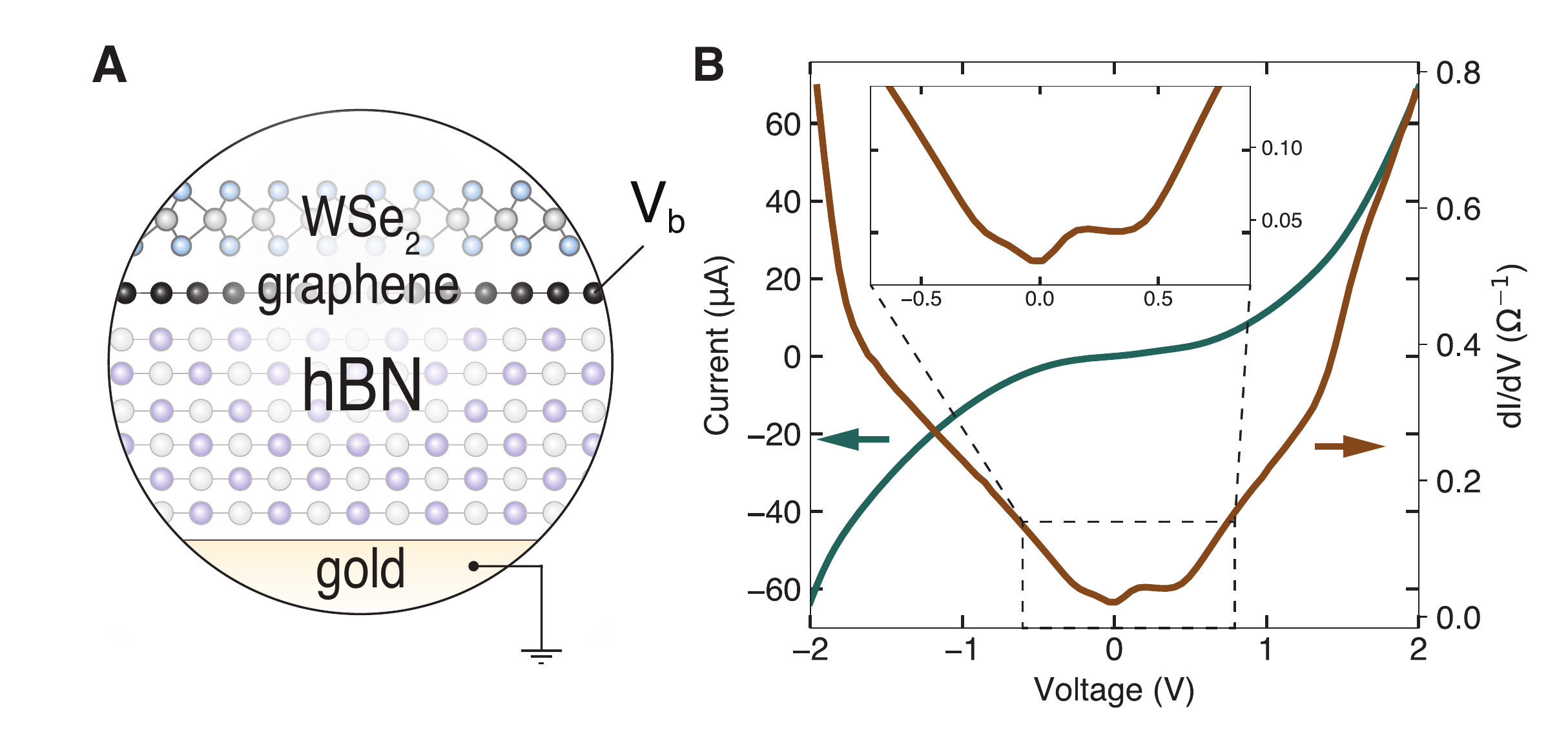}
		\caption{ \textbf{Electrical characterization.} Device schematic of a TMD coupled tunneling junction. Current and conductance (dI/dV) measurements as a function of \vbias~.} 
		\label{figs0}
		
	\end{figure}
	
\end{center}

We electrically characterize the device by applying a bias voltage (\vbias) between the gold and graphene electrodes as indicated in Fig.~\ref{figs0}A. The current and conductance measurements are shown in Fig.~\ref{figs0}B. The triple minima in the conductance curve around 0 V are associated with the intrinsic graphene doping and the phonon induced tunneling that governs the transport of such devices, as it has been investigated in previous studies \cite{Zhang2008, Vdovin2016, Parzefall2019}. This indicates that tunneling indeed occurs between graphene and gold and is phonon-mediated. Moreover, for \vbias~$<$1.5 V the conductance shows a linear behavior. In this low voltage regime compared to the \hBN barrier (\SI{5}{eV}) the conductance curve is associated to the linear density of states of graphene. Interestingly, around \vbias = \SI{1.5}{V} and \SI{-1.7}{V}, the slope of the conductance increases. This behavior is not observed in devices without a \wse2 and can be associated with an increased rate due to interaction of the tunneling electrons with excitonic states.

\pagebreak
\subsection*{Exciton generation in both polarities}

\begin{center}
	\begin{figure}[h!]\center
		\includegraphics[width=0.8\textwidth]{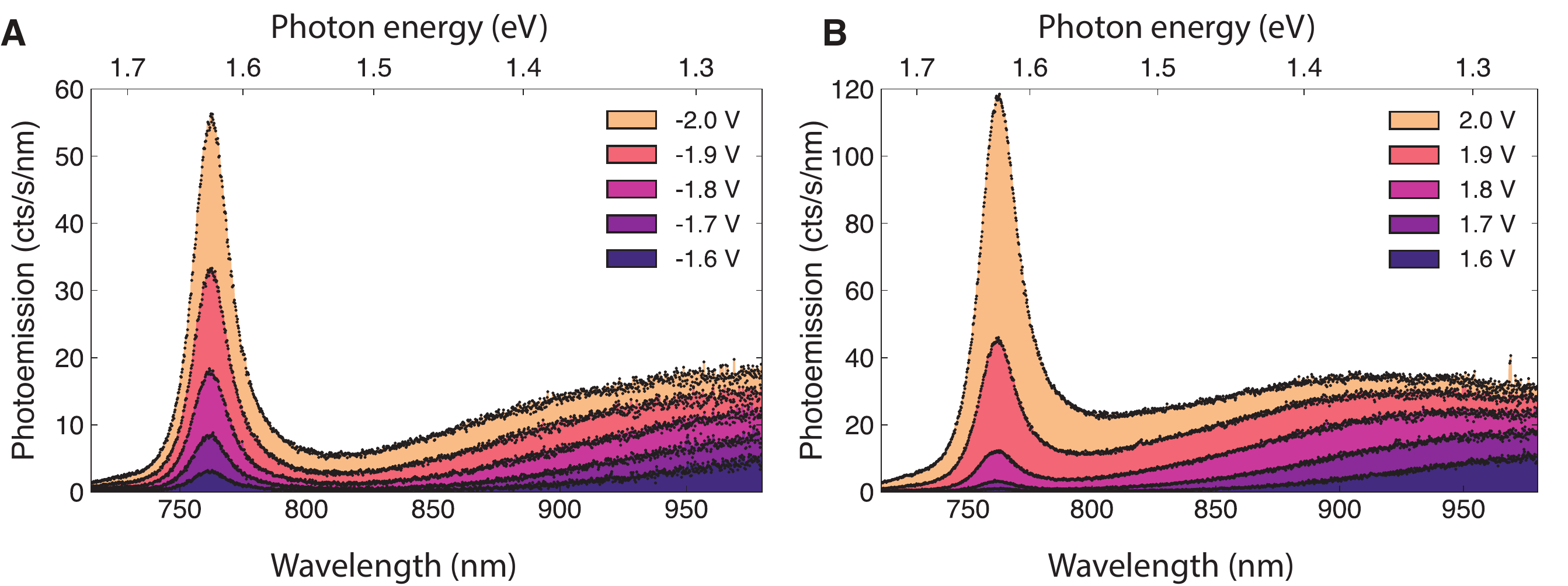}
		\caption{\textbf{Spectral symmetry with bias voltage.} (a,b) Photoemission spectra from the TMDC coupled tunneling junction presented in Fig.~2A in the main text for (a) positive bias voltage and (b) negative bias voltage. Similar spectral characteristics are observed. This observation further supports the argument that there excitons are not created by charge injection as such a process would be polarity dependent.} 
		\label{figs1}
	\end{figure}
	
\end{center}

As discussed in the main text the exciton generation process occurs for both bias polarities. Figure~\ref{figs1} shows spectra for both polarities where the narrow emission at exciton wavelengths is This indicates excitons are not generated by electron-hole injection since such a process, if possible, would be strictly polarity dependent. 

\newpage
\subsection*{TMD coupled device with 5 nm spacer hBN}

\begin{center}
	\begin{figure}[h!]\center
		\includegraphics[width=0.9\textwidth]{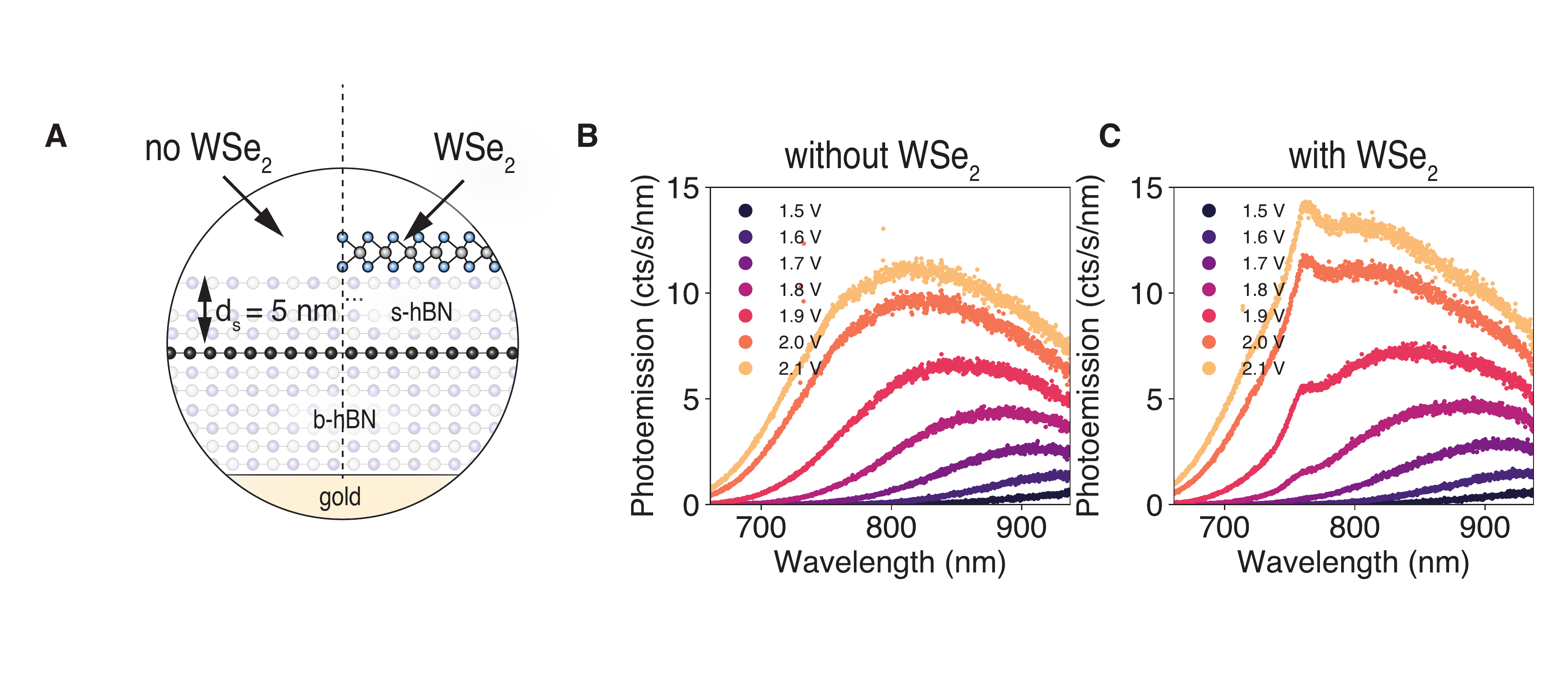}
		\caption{\textbf{TMD coupled device with 5 nm spacer hBN.} (a) Device schematic. The \wse2 is separated from the tunneling junction by a spacer \hBN with 5 nm thickness. The \wse2 flake is stacked on top in a way that two areas are formed: with and without \wse2. (b,c) Photoemission spectra from the two areas (b) without \wse2 and (c) with \wse2. A small contribution in excitonic wavelengths is still observed even though the \wse2 flake is 5 nm away from the tunneling junction.} 
		\label{figs2}
	\end{figure}
	
\end{center}